\documentclass{ifacconf}
\makeatletter
\let\old@ssect\@ssect 
\makeatother
\usepackage[hidelinks]{hyperref}
\makeatletter
\def\@ssect#1#2#3#4#5#6{%
	\NR@gettitle{#6}
	\old@ssect{#1}{#2}{#3}{#4}{#5}{#6}
}
\makeatother
\usepackage{graphicx}      
\usepackage{natbib}        

\usepackage{nccmath}

\usepackage{amssymb}    
\usepackage{mathtools}  
\usepackage{pgfplots}
\usepackage{tikz}
\usepgfplotslibrary{external}
\usetikzlibrary{shapes}
\usetikzlibrary{positioning}
\usetikzlibrary{arrows,decorations.markings}
\usepackage{bm}

\usepackage{caption}
\usepackage{subcaption}
\usepackage{siunitx}

\usepackage{pgfplots}
\usepackage{tikz}
\usepgfplotslibrary{external}
\usetikzlibrary{shapes}
\usetikzlibrary{positioning}
\usetikzlibrary{arrows,decorations.markings}
\usetikzlibrary{spy}

\newtheorem{theorem}{Theorem}

\newtheorem{lemma}{Lemma}
\newenvironment{proof}{Proof:}{\hfill$\square$\\}

\begin{document}
	{\onecolumn \begin{center} Accepted for publication at the IFAC Conference on Networked Systems (NecSys) 2022\end{center}
	
	\noindent\fbox{%
		\parbox{\textwidth}{%
				©2022 the authors. This work has been accepted to IFAC for publication under a Creative Commons Licence CC-BY-NC-ND.
		}%
	}
	\newpage
\begin{frontmatter}

\title{Event-triggered and distributed model predictive control for guaranteed collision avoidance in UAV swarms \thanksref{footnoteinfo}} 

\thanks[footnoteinfo]{This work was supported in part by the German Research Foundation (DFG) within the priority program 1914 (grant TR 1433/1-2).}

\author[First]{Alexander Gr\"afe} 
\author[First]{Joram Eickhoff} 
\author[First]{Sebastian Trimpe}

\address[First]{Institute for Data Science in Mechanical Engineering (DSME), RWTH Aachen University, 52068 Aachen, Germany (e-mail: alexander.graefe@dsme.rwth-aachen.de).}

\begin{abstract}                
Distributed model predictive control (DMPC) is often used to tackle path planning for unmanned aerial vehicle (UAV) swarms. However, it requires considerable computations on-board the UAV, leading to increased weight and power consumption. In this work, we propose to offload path planning computations to multiple ground-based computation units. As simultaneously communicating and recomputing all trajectories is not feasible for a large swarm with tight timing requirements, we develop a novel event-triggered DMPC that selects a subset of most relevant UAV trajectories to be replanned. The resulting architecture reduces UAV weight and power consumption, while the active redundancy  provides robustness against computation unit failures. Moreover, the DMPC guarantees feasible and collision-free trajectories for UAVs with linear dynamics. In simulations, we demonstrate that our method can reliably plan trajectories, while saving \SI{60}{\percent} of network traffic and required computational power. Hardware-in-the-loop experiments show that it is suitable to control real quadcopter swarms.
\end{abstract}

\begin{keyword}
Distributed constrained control and MPC, Event-triggered and self-triggered control, Multi-agent systems 
\end{keyword}

\end{frontmatter}

\newcommand{\fakepar}[1]{\vspace{1mm}\noindent\textit{#1.}}
\newcommand{\capt}[1]{\mdseries{\emph{#1}}}
\newcommand*{\matlab}{\textit{Matlab}}
\newcommand*{\simulink}{\textit{Simulink}}
\newcommand*{\purepursuit}{\textit{Pure Pursuit}}
\newcommand*{\carCup}[0]{\textit{Carolo-Cup}~}
\newcommand*{\ros}[0]{\textit{ROS2}~}
\newcommand*{\psaf}[0]{Projektseminar Autonomes Fahren~}
\newcommand*{\darpa}[0]{\textit{DARPA}~}
\newcommand*{\opencv}[1]{\textit{OpenCV}}

\newcommand*{\jt}{\textit{Jetson TX2}}
\newcommand*{\jpi}{\textit{SDK Manager}}

\newcommand*{\phil}{\ensuremath{\varphi_\textrm{L}}}
\newcommand*{\philmax}{\ensuremath{\varphi_\textrm{L,max}}}
\newcommand*{\dphilmax}{\ensuremath{\dot{\varphi_\textrm{L,max}}}}

\newcommand*{\mat}[1]{{\ensuremath{\mathrm{\textbf{#1}}}}}
\newcommand*{\ma}[1]{{\ensuremath{\boldsymbol{\mathrm{#1}}}}}
\newcommand*{\mas}[1]{\ensuremath{\boldsymbol{#1}}}
\newcommand*{\ve}[1]{{\ensuremath{\boldsymbol{#1}}}}
\newcommand*{\ves}[1]{\ensuremath{\boldsymbol{\mathrm{#1}}}}

\newcommand*{\AP}{\ensuremath{\mathrm{AP}}}
\newcommand*{\doti}{\ensuremath{(i)^\cdot}}

\newcommand*{\inprod}[2]{\ensuremath{\langle #1,\,#2 \rangle}}


\newcommand*{\ud}{\ensuremath{\mathrm{d}}}

\newcommand*{\tn}[1]{\textnormal{#1}}

\newcommand*{\mrm}[1]{\ensuremath{\mathrm{#1}}}

\newcommand*{\transp}{\ensuremath{\mathrm{T}}}

\newcommand*{\rang}{\ensuremath{\operatorname{rg}}}

\newcommand*{\grpsb}[2]{\ensuremath{\left(#1\right)_{#2}}}
\newcommand*{\grprsb}[2]{\ensuremath{\left(#1\right)_{\mathrm{#2}}}}

\newcommand*{\normd}[2]{\ensuremath{\frac{\mathrm{d}#1}{\mathrm{d}#2}}}
\newcommand*{\normdat}[3]{\ensuremath{\left.\frac{\mathrm{d} #1}{\mathrm{d} #2}\right|_{#3}}}

\newcommand*{\matd}[2]{\ensuremath{\frac{\mathrm{D} #1}{\mathrm{D} #2}}}
\newcommand*{\matdat}[3]{\ensuremath{\left.\frac{\mathrm{D} #1}{\mathrm{D} #2}\right|_{#3}}}

\newcommand*{\partiald}[2]{\ensuremath{\frac{\partial #1}{\partial #2}}}
\newcommand*{\partialdat}[3]{\ensuremath{\left.\frac{\partial #1}{\partial #2}\right|_{#3}}}

\newcommand*{\FT}[1]{\ensuremath{\mathfrak{F}\left\{#1\right\}}}
\newcommand*{\FTabs}[1]{\ensuremath{\left|\mathfrak{F}\left\{#1\right\}\right|}}
\newcommand*{\IFT}[1]{\ensuremath{\mathfrak{F}^{-1}\left\{#1\right\}}}
\newcommand*{\DFT}[1]{\ensuremath{\mathrm{DFT}\left\{#1\right\}}}
\newcommand*{\DFTabs}[1]{\ensuremath{\left|\mathrm{DFT}\left\{#1\right\}\right|}}
\newcommand*{\Laplace}[1]{\ensuremath{\mathfrak{L}\left(#1\right)}}
\newcommand*{\InvLaplace}[1]{\ensuremath{\mathfrak{L^{-1}}\left(#1\right)}}
\newcommand*{\invtrans}{\ensuremath{\quad\bullet\!\!-\!\!\!-\!\!\circ\quad}}
\newcommand*{\trans}{\ensuremath{\quad\circ\!\!-\!\!\!-\!\!\bullet\quad}}

\newcommand*{\textcompstdfont}[1]{{\fontfamily{cmr} \fontseries{m} \fontshape{n} \selectfont #1}}

\newcommand*{\mlfct}[1]{\texttt{#1}}

\newcommand*{\UL}[2]{#1_\mathrm{#2}}
\newcommand*{\ULi}[2]{#1_{#2}}
\newcommand*{\dy}[0]{\dot{y}}
\newcommand*{\ddy}[0]{\ddot{y}}
\newcommand*{\receivedPower}[0]{\frac{A}{y^2+h^2}}
\newcommand*{\dD}[0]{\dot{D}}
\newcommand*{\tf}[0]{\UL{t}{f}}
\newcommand*{\umax}[0]{\UL{u}{max}}

\newcommand*{\x}[1]{\UL{x}{#1}}
\newcommand*{\xv}[0]{\ve{x}}
\newcommand*{\la}[1]{\UL{\lambda}{#1}}
\newcommand*{\ua}[0]{\UL{u}{+}}
\newcommand*{\ub}[0]{\UL{u}{-}}
\newcommand*{\Ht}[1]{\tilde{H}(#1)}
\newcommand*{\Dp}[0]{\UL{D}{p}}
\newcommand*{\Dm}[0]{\overline{\dot{D}}}
\newcommand*{\fdynamic}[0]{\UL{\ve{f}}{dynamic}(\ve{y}, u)}
\newcommand*{\fE}[0]{\UL{f}{E}(\ve{y}, u)}
\newcommand*{\fD}[0]{\UL{f}{D}(\ve{y}, u)}
\newcommand*{\vl}[0]{\UL{v}{l}}
\newcommand{\vex}[0]{\ve{x}}
\renewcommand*{\of}[1]{\left(#1\right)}
\newcommand*{\vexof}[1]{\vex\of{#1}}
\newcommand*{\vexul}[1]{\UL{\vex}{#1}}
\newcommand*{\vexulof}[2]{\UL{\vex}{#1}\of{#2}}
\newcommand*{\vexuli}[1]{\ULi{\vex}{#1}}
\newcommand*{\hatvexuli}[1]{\ULi{\hat{\vex}}{#1}}
\newcommand*{\vexuliof}[2]{\ULi{\vex}{#1}\of{#2}}
\newcommand*{\hatvexuliof}[2]{\ULi{\hat{\vex}}{#1}\of{#2}}
\newcommand*{\pcom}[0]{{P\ul{com}}}
\newcommand*{\pcomhat}[0]{\hat{P}\ul{com}{}}
\newcommand{\weight}[2]{w\uliof{#1}{#2}}
\newcommand{\history}[0]{\mathcal{E}}
\newcommand{\weighthat}[2]{\hat{w}\uliof{#1}{#2}}

\newcommand*{\veerr}[0]{\ve{e}}
\renewcommand*{\ul}[1]{_{\mathrm{#1}} }
\newcommand*{\uliof}[2]{_{#1}{\of{#2}}}
\newcommand*{\uli}[1]{_{#1} }
\newcommand*{\koop}[0]{\mathcal{K}}
\newcommand*{\koopt}[0]{\mathcal{K}^t}
\newcommand*{\gof}[1]{g\of{#1}}
\newcommand*{\gul}[1]{\ULi{g}{#1}}
\newcommand*{\gofxof}[1]{\gof{\vexof{#1}}}
\newcommand*{\phivexof}[1]{\phi\of{\vexof{#1}}}
\newcommand*{\veg}[0]{\ve{g}}
\newcommand*{\vegof}[1]{\ve{g}\of{#1}}
\newcommand*{\vegulof}[2]{\ve{g}_{\mathrm{#1}}\of{#2}}
\newcommand*{\vey}[0]{\ve{y}}
\newcommand*{\veyof}[1]{\vey\of{#1}}
\newcommand*{\veyul}[1]{\UL{\vey}{#1}}
\newcommand*{\tzero}[0]{\UL{t}{0}}
\newcommand*{\vepsi}[0]{\ve{\psi}}
\newcommand*{\vephi}[0]{\ve{\phi}}
\newcommand*{\veu}[0]{\ve{u}}
\newcommand*{\veuof}[1]{\veu\of{#1}}
\newcommand*{\veuul}[1]{\UL{\veu}{#1}}
\newcommand*{\vev}[0]{\ve{v}}
\newcommand*{\vevof}[1]{\vev\of{#1}}
\newcommand*{\vevul}[1]{\UL{\vev}{#1}}
\newcommand*{\xul}[1]{\ULi{x}{#1}}
\newcommand*{\aul}[1]{\UL{a}{#1}}
\newcommand*{\kul}[1]{\UL{k}{#1}}
\newcommand*{\uul}[1]{\ULi{u}{#1}}

\newcommand*{\maAuli}[1]{\ULi{\ma{A}}{#1}}
\newcommand*{\veeps}[0]{\ve{\epsilon}}
\newcommand*{\twonorm}[1]{||#1||_\mathrm{2}}

\newcommand*{\lambdamaxof}[1]{\UL{\lambda}{max}\of{#1}}
\newcommand*{\lambdaminof}[1]{\UL{\lambda}{min}\of{#1}}
\newcommand*{\B}[0]{Method B\:}

\newcommand*{\tensorflowprob}{\textit{Tensorflow-Probability}~}

\newcommand*{\EX}[0]{\mathbb{E}}
\newcommand*{\tr}[0]{\mathrm{Trace}}
\newcommand*{\var }[0]{\mathrm{Var}}
\newcommand*{\given}[0]{\;\middle|\;}
\newcommand*{\gpess}[0]{g\ul{pess}{}}
\newcommand*{\gopt}[0]{g\ul{opt}{}}
\newcommand*{\tveerr}[0]{\tilde{\veerr}}
\newcommand*{\ofs}[1]{\left[#1\right]}

\newcommand*{\matA}[1]{\ma{A}\uli{#1}}
\newcommand*{\matGamma}[0]{\ma{\Gamma}}
\newcommand*{\matGammaOpt}[0]{\ma{\Gamma}\ul{opt}}

\newcommand*{\matQ}[1]{{\ma{Q}\uli{#1}}}
\newcommand*{\matP}[1]{\ma{P}\uli{#1}}
\newcommand*{\eps}[2]{\epsilon\uli{#1}\of{#2}}
\newcommand*{\alp}[2]{\alpha\uli{#1}\of{#2}}
\newcommand*{\pfail}[0]{P\ul{loss}}
\newcommand{\priority}[2]{g\uli{#1}\of{\errorest{#1#1}{#2}}}
\newcommand*{\diag}[0]{\text{diag}}
\newcommand*{\tp}[1]{\of{#1}}
\newcommand*{\state}[2]{\vex\uli{#1}\tp{#2}}
\newcommand*{\stateest}[2]{\hat{\vex}\uli{#1}\tp{#2}}
\newcommand*{\noise}[2]{\ve{w}\uli{#1}\tp{#2}}
\newcommand*{\processnoise}[2]{\ve{v}\uli{#1}\tp{#2}}
\newcommand*{\errart}[2]{\ve{\tilde{e}}\uli{#1}\tp{#2}}
\newcommand*{\errorest}[2]{\ve{e}\uli{#1}\tp{#2}}
\newcommand*{\barerrorest}[2]{\ve{\bar{e}}\uli{#1}\tp{#2}}
\newcommand*{\errorestmeas}[2]{\ve{e}\uli{#1, meas}\tp{#2}}
\newcommand*{\erroragents}[2]{\hat{\ve{e}}\uli{#1}\tp{#2}}
\newcommand*{\erroragentstotal}[2]{\underline{\hat{\ve{e}}}\uli{#1}\tp{#2}}
\newcommand*{\erroresttotal}[2]{\underline{\ve{e}}\uliof{#1}{#2}}
\newcommand*{\matB}[1]{\ma{B}\uli{#1}}
\newcommand*{\matV}[1]{\ma{V}\uli{#1}}
\newcommand*{\matVti}[1]{\ma{\tilde{V}}\uli{#1}}
\newcommand*{\matW}[1]{\ma{W}\uli{#1}}
\newcommand*{\matAT}[1]{\left(\ma{A}\uli{#1}^\transp\right)}
\newcommand*{\matF}[1]{\ma{F}\uli{#1}}
\newcommand*{\rec}[2]{\gamma\uli{#1}\tp{#2}}
\newcommand*{\Cov}{\mathrm{Cov}}
\newcommand*{\matAti}[1]{\tilde{\ma{A}}\uli{#1}}

\newcommand{\io}[1]{\Gamma\of{#1}}
\newcommand{\oi}[1]{\Lambda\of{#1}}

\newcommand{\matAd}[1]{\tilde{\mat{A}}\ul{d, \mathrm{#1}}}
\newcommand{\ri}[1]{{r\uli{#1}}}
\newcommand{\di}[1]{{d\uli{#1}}}
\newcommand{\nii}[1]{{n\uli{#1}}}
\newcommand{\deltai}[1]{\Delta\uli{#1}}
\newcommand{\betai}[1]{\beta\uli{#1}}
\newcommand{\alphai}[1]{\alpha\uli{#1}}
\newcommand{\zetai}[1]{\zeta\uli{#1}}
\newcommand{\lmax}[1]{\lambda\ul{max}\of{#1}}
\newcommand{\lmin}[1]{\lambda\ul{min}\of{#1}}
\newcommand{\unknown}[0]{\left(\cdot\right)}
\newcommand*{\kt}[0]{\tilde{k}}
\newcommand*{\nz}[0]{\ve{n}\ul{0}}
\newcommand*{\norm}[2]{||#1||_{#2}}
\newcommand*{\diff}[0]{\text{d}}

\newcommand{\dalpha}[0]{\delta\alpha}
\newcommand{\setS}[0]{\mathcal{S}}
\newcommand{\cone}[0]{c\ul{1}\of{\setS}}
\newcommand{\ctwo}[0]{c\ul{2}\of{\setS}}

\newcommand{\pos}[0]{\ve{p}}
\newcommand{\speed}[0]{\ve{v}}
\newcommand{\acc}[0]{\ve{a}}

\newcommand{\posset}[1]{\pos\uli{#1, \mathrm{set}}}
\newcommand{\speedset}[1]{\speed\uli{#1, \mathrm{set}}}
\newcommand{\accset}[1]{\acc\uli{#1, \mathrm{set}}}
\newcommand{\vexset}[1]{\vex\uli{#1, \mathrm{set}}}
\renewcommand{\posset}[1]{\pos\uli{#1}}
\renewcommand{\speedset}[1]{\speed\uli{#1}}
\renewcommand{\accset}[1]{\acc\uli{#1}}
\renewcommand{\vexset}[1]{\vex\uli{#1}}
\newcommand{\stimeobj}[0]{T\ul{o}}
\newcommand{\stimes}[0]{T\ul{s}}
\newcommand{\stimecons}[0]{T\ul{b}}
\newcommand{\stimecol}[0]{T\ul{c}}
\newcommand{\dxmax}[2]{\Delta\vexset{#1}{}\ul{max}\of{#2}}
\newcommand{\dpmax}[2]{\Delta\posset{#1}{}\ul{,max}\of{#2}}

\newcommand{\cu}{CU}
\newcommand{\cus}{CUs}
\newcommand{\artinput}{\ve{u}}
\newcommand{\artinputset}{\mathcal{u}}
\section{Introduction}
In the recent years, unmanned aerial vehicle (UAV) swarms have become larger and larger, reaching sizes from 50 up to 100 UAVs (\cite{Crazyswarm}). To find trajectories that do not lead to colliding UAVs for such big swarms poses a challenge for path planning algorithms. The planning usually requires a lot of computational power, which can either be centralized or distributed among the UAVs. In the first case, a stationary computer transmits the trajectories to the UAVs over a wireless network. In the latter, the UAVs calculate their paths on-board and have to exchange information among themselves over a wireless network.
Central model predictive control (MPC) approaches like \cite{Augugliaro2012} need no powerful processor on board of the UAV. This decreases weight and thus energy consumption of the UAV. However, they do not scale well with an increasing number of UAVs (\cite{Luis2019}) and expose the system to the potential risk of a whole system failure if the central computer fails (\cite{alzain2012cloud}). Distributed model predictive control (DMPC) based approaches have become increasingly popular (\cite{Zhou2017}, \cite{Luis2019}, \cite{Luis2020} \cite{Park2021}). First, they do not require any central node since all computations are done on the UAVs. Second, they scale well with increasing number of UAVs. However, the on-board processing needs computational capacity and thus increases weight, leading to a higher energy consumption of the UAVs. 

This work uses the ``best of both'' and combines centralized and distributed features based on a novel event-triggered DMPC (ET-DMPC) design.
 Our approach consists of multiple stationary computation units (CU, Figure \ref{fig:overview}). Every \cu\ is placed on the ground and has enough power to replan the trajectory of one UAV at a time. It shares the resulting trajectory with other \cus\ and the UAVs over a wireless network, leveraging recent advances in low-power wireless technology (\cite{PredictiveTriggering}). The UAVs then follow their respective
 trajectories. However, it is not viable to provide one CU for each UAV in the swarm, which makes the system more expensive and may exceed the available network bandwidth. We thus propose to use fewer \cus\ than UAVs and present an event-triggering (ET) mechanism that enables trajectory replanning for only a subset of UAVs at a time.
 
This approach has benefits over purely centralized or distributed approaches. First, the UAVs need less on-board computational power, which saves weight and reduces energy consumption. Second, in contrast to standard centralized approaches, it is more robust against failures, because a failure of one CU does not lead to a whole system failure, as the remaining CUs in combination with ET can still plan the paths of the UAVs.
The presented ET-DMPC uses features from existing DMPCs and scales well with the size of the swarm.
In summary, this paper makes the following contributions:
\begin{itemize}
	\item A novel distributed computing method for path planning of large UAV swarms which uses ET for improved resource efficiency;
	\item a new ET-DMPC formulation based on existing DMPC approaches (\cite{Luis2019,Luis2020, Park2021});
	\item formal proofs of recursive feasibility and collision free trajectories for UAVs with linear dynamics;
	\item an experimental evaluation of the effect that the ET has on the overall performance\footnote{A video of the method controlling a real quadcopter swarm can be found in \url{https://youtu.be/2UzqOnUJQCA}.}. 
\end{itemize}  

\begin{figure}
	\centering
	\includegraphics[width=0.9\linewidth]{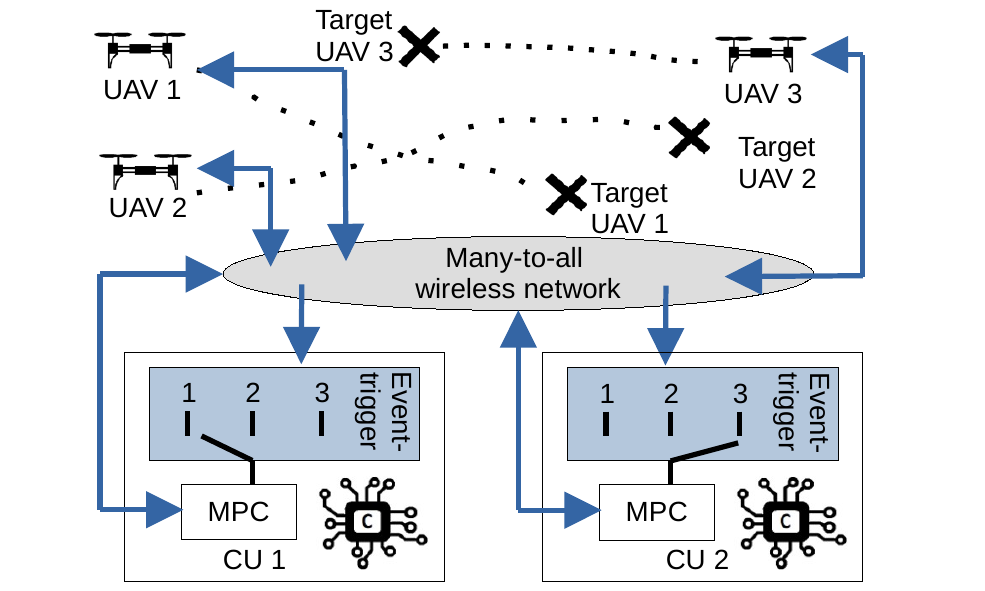}
	\caption{Proposed event-triggered distributed Model Predictive Control (ET-DMPC) setup for path planning of UAV swarms. \capt{The computation units (CU) solve an MPC problem. They transmit the solution to the UAVs. An event-trigger decides which UAV's paths are replanned.}}
	\label{fig:overview}
\end{figure}
\section{Related Work}
Many DMPC based trajectory planners  (\cite{Zhou2017, Cai2018, Luis2019, Luis2020, Park2021}) have the same communication structure but differ in their use of constraints and UAV models. In this structure, every UAV $i$ has knowledge about the last planned path of all other UAVs. It uses them to creates pairwise anti-collision constraints between them and the new planned path of itself. These constraints are used in an optimization problem, whose solution is the new planned path. In a next step, all UAVs share their calculated paths over a communication network with all other UAVs. We build on the same structure and leverage features of prior work such as the communication structure and DMPC formulation, however, we add \cus\ with ET to save resources and combine existing techniques to guarantee recursive feasibility and collisions free trajectories. 

Although our ET mechanism can be used to adapt existing DMPC approaches, we will combine constraints and UAV models from different works to create our own DMPC algorithm, which guarantees feasibility and collision-free trajectories.
In this work, we will use the approach of \cite{Luis2019} and \cite{Luis2020} to use a linear state space model and linear anti-collision constraints to generate a quadratic-programming problem (QP). However, the anti-collision constraints presented in \cite{Luis2019} do not guarantee collision free trajectories, because the pairwise constraints have common intersections (see Figure \ref{fig:constrComp}). 
One method, which guarantees collision free trajectories are Buffered Voronoi Cells (BVC). They restrict the positions of the agents to be in separated sets (\cite{Zhou2017}). However, for large swarms, these sets are small and lead to poor performance (\cite{Luis2020}).
To overcome this problem, \cite{Park2021} use time-variant constraints based on the convex hull property of Bernstein polynomials. These constraints ensure that there are no collisions along the entire continuous trajectory. \cite{TVBVC} present a similar approach, where BVC are varied at every sample point of the optimization problem. We build on their idea of time-variant BVC (TV-BVC) in this paper. 
Additional, we will use the idea of \cite{Park2021} to introduce additional constraints to ensure recursive feasibility.
Concurrent with this work, \cite{Chen2022} have developed a similar strategy using TV-BVC and additional constraints to ensure recursive feasibility. However, the approach does not use ET to lower the communication bandwidth and calculates the solution of the DMPC on the UAVs. 

\cite{Cai2018} present an ET-DMPC for quadcopter swarms. The DMPC uses nonlinear dynamics but does not guarantee feasibility and collision free trajectories. The ET has multiple triggering conditions and triggers recalculation if at least one of them is fulfilled. However, none of the conditions takes the limited computation power directly into account. This approach is thus not suitable for our setup, because the \cus\ can only replan a fixed number of UAVs at once. 

  
\section{Problem Setting}
As a key conceptual difference to the aforementioned works, we introduce an architecture with $M$ \cus\ to control $N\geq M$ UAVs (see Figure \ref{fig:overview}).
In the following, the task of the UAVs is to fly to target positions without colliding with each other. While the UAVs' on-board computers do not have enough computational power to solve a trajectory planning problem, every \cu\ has enough capacity to solve it for one UAV at a time. In order to share information, UAVs and \cus\ use a wireless network. In this work, we assume a perfect connection without any message loss. However, we assume that the network has limited bandwidth and some delay.

The problem considered in this work splits into three different subproblems.
First, we need to design a \textbf{communication architecture} to facilitate robust data exchange between multiple stationary CUs and mobile UAVs.
Second, we need to formulate the \textbf{DMPC} that runs on the \cus. In this paper, as done in related works (\cite{Zhou2017,Luis2019, Luis2020, Wang2021}), we assume that the dynamics of the UAVs can be modeled by a linear system. We also assume that they are able to hover, e.g. like multicopters or balloon-based robots. The overall dynamics are described as
\begin{equation}
	\label{eq:dynamics}
	\begin{split}
	\dot{\vex}\uli{i} &= 
	\begin{bmatrix}
		\dot{\pos}\uli{i}\\
		\dot{\speed}\uli{i}\\
		\dot{\vey}\uli{i}
	\end{bmatrix} = \begin{bmatrix}\ma{0}&\ma{I}\uli{3}&\ma{0}\\
	\ma{0}&\ma{A}\uli{22, i}& \ma{A}\uli{23, i}\\
	\ma{0}&\ma{A}\uli{32, i}& \ma{A}\uli{33, i}
\end{bmatrix}\begin{bmatrix}
\pos\uli{i}\\
\speed\uli{i}\\
\vey\uli{i}
\end{bmatrix} +\begin{bmatrix}
\ma{0}\\
\ma{B}\uli{2, i}\\
\ma{B}\uli{3, i}
\end{bmatrix}\veu\uli{i}\\
&=\matA{i}\vex\uli{i} + \matB{i}\veu\uli{i},
\end{split}
\end{equation}
where $\pos\uli{i}$ and $\speed\uli{i}$ describe the position and speed of the UAV, $\vey\uli{i}$ summarizes the rest of the statespace of UAV $i$ (e.g. acceleration, jerk). $\veu\uli{i, \mathrm{min}}\leq\veu\uli{i}\leq\veu\uli{i, \mathrm{max}}$ is the control input of the system.
Third, the swarm and the \cus\ need to decide which UAV's trajectory gets replanned using \textbf{ET}. Because the system has more UAVs than \cus, the ET has to take the exact number of \cus\ directly into account. 

The combination of the network, DMPC and ET should steer the UAVs to their targets $\vex\uli{i,\mathrm{target}}=[\pos\uli{i, \mathrm{target}},\ve{0}, \ve{0}]$. To maintain safety, they have to guarantee that the UAVs do not collide. For collision-free trajectories, we will require that the positions are never closer than $r\ul{min}$,
\begin{equation}
	\label{eq:col}
	\begin{split}
	\twonorm{\ma{\Theta}^{-1}\left[\pos\uli{i}\of{t}-\pos\uli{j}\of{t}\right]} \geq r\ul{min},\\ \forall t>0~\forall i,j\in\left\{1,...,N\right\}, i\neq j.
	\end{split}
\end{equation} Because some types of UAV need to maintain a higher distance in one direction, e.g. due to downwash, the coordinate system is scaled using a symmetric positive definite matrix $\ma{\Theta}$ (\cite{Luis2019, Luis2020, Park2021}).
Additionally, the whole system has to guarantee that the DMPC problem is feasible at all time. Otherwise, the UAVs might collide because the algorithm fails to find a trajectory.

\section{Event-triggered Distributed Model Predictive Control}
\begin{figure}
	\centering
	\subcaptionbox{Linearized constraints (\cite{Luis2019})}[.2\textwidth]{\includegraphics[width=.2\textwidth]{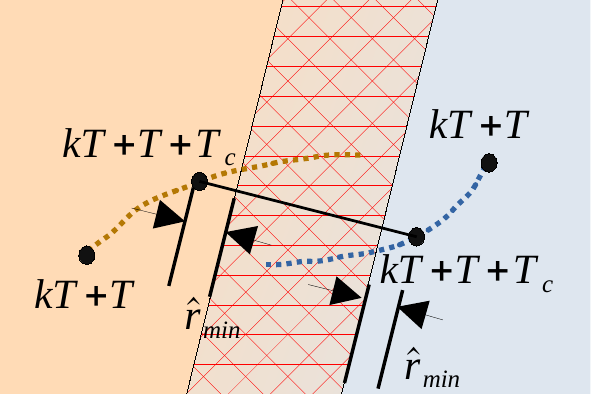}}
\subcaptionbox{TV-BVC (\cite{TVBVC})}[.2\textwidth]{\includegraphics[width=.2\textwidth]{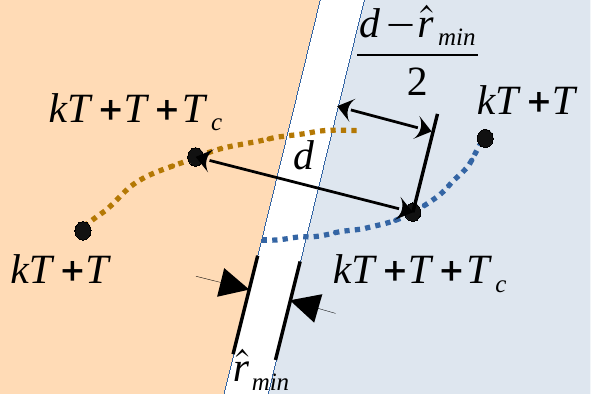}}
\caption{Comparison of different anti-collision constraints for the trajectory at time $kT+T+\stimecol$. \capt{The colored areas specify the corresponding constraint of the trajectory with the same color. (a): Linearization of the constraints. There exists a set, where both trajectories can lie in. 
 (b): Time variant BVC (TV-BVC). The generated constraints restrict the allowed set of trajectories to separated sets. Here $d=||\ve{n}\uli{ij}\of{\stimecol+2T|(k-1)T}||_2$. 
}}
\label{fig:constrComp}
\end{figure}
In this section, we will now solve the three subproblems that we have defined in the section above.\\
{\bf Communication architecture.} The choice of the communication architecture is mainly influenced by related works (\cite{Zhou2017, Cai2018, Luis2019, Luis2020, Park2021}). To simplify the design of the DMPC and the ET, the algorithms should not have to deal with different timings and information states of the agents. Therefore, we choose a round based many-to-all architecture. During every round, it distributes all messages sent to all agents in the network. This synchronizes the agents, i.e they send/receive at the same time, and all agents have the same knowledge about the data sent in the network. From an implementation based point of view, this choice is reasonable too. There exist real world implementations of round based many-to-all architectures like Glossy (\cite{Ferrari2011}) or Mixer (\cite{Mixer}), which have been shown to meet the strict real-time and reliability requirements needed for fast control (\cite{FeedbackControlGoesWireless, PredictiveTriggering}). 

During the communication round, the UAVs send their current state and the \cus\ share their results of the path planning, which the corresponding UAVs and all \cus\ save in a buffer. 
After the communication round, the \cus\ decide which \cu\ replans the path of which UAV using ET. Because of the many-to-all protocol, every \cu\ has the same information and thus knows which UAV will be selected by the other \cus. This avoids the case of two \cus\ replanning the trajectory of the same UAV. Afterwards, each unit replans the trajectory of its chosen UAV. The maximum duration for the scheduling and replanning $T\ul{calc}$ is constant for every agent and time. A new communication round (length $T\ul{com}$) starts afterwards. The combined duration between the starts of two rounds is $T=T\ul{calc}+T\ul{com}$.

{\bf ET-DMPC.} The event trigger decides if during the $k$-th round, \cu\ $q$ replans the path of UAV $i$ ($\gamma\uli{q i}\of{k}=1$) or not ($\gamma\uli{q i}\of{k}=0$). Every \cu\ can only replan the path of one UAV in one round and every \cu\ should replan the path of a different UAV. 
 In this work, we use a priority based trigger rule (PBT) (\cite{PredictiveTriggering}), which we present at the end of this section, because it uses features from the DMPC, which we present next.

The control input for each UAV is piecewise-constant with sampling time $\stimes$ and horizon $h\ul{s}$
\begin{equation}
	\label{eq:discretisation}
		\veu\uli{i}\of{\tau+T|kT} = \sum_{\kappa=0}^{h\ul{s}-1}\Gamma\ul{\stimes}\of{\tau-\kappa\stimes}\veu\uli{i, \kappa|k},
\end{equation}
where $\Gamma\ul{\stimes}\of{t}$ is equal to $1$ for $0\leq t< \stimes$ and $0$ otherwise, and $\veu\uli{i, \kappa|k}$ are the time-discrete inputs. Additionally, we require $\tfrac{T}{\stimes} = \tfrac{h\ul{s}}{H}\in\mathbb{N}$ to ensure recursive feasibility of the following optimization problem with prediction horizon $H$ (\cite{Park2021}). When a \cu\ recalculates the trajectory of UAV $i$ at time $kT$, it solves the following optimization problem:
\begin{subequations}
	\label{eq:opt}
\begin{align}
	 \label{eq:objectfunc}
	\min_{\artinput\uli{i, \cdot|k} }\sum_{\kappa=0}^{h\ul{o}}\big[&||\vexset{i}\of{\kappa\stimeobj+T|kT}-\vex\uli{i, \mathrm{target}}||_{\ma{Q}\uli{i}}^2 \\&+ ||\veu\uli{i}\of{\kappa\stimeobj+T|kT}||_{\ma{R}\uli{i}}^2\big]\nonumber
\end{align}
\begin{align}
	\text{s.t.~}&\vex\uli{i}\of{\kappa\stimes+T+\delta|kT}\label{eq:dynconst}\\
	& = e^{\ma{A}\uli{i}(\kappa\stimes+\delta)}\vex\uli{i
	}\of{2T|(k-1)T}+\int_{0}^{\delta}e^{\ma{A}\uli{i}\tau}\ma{B}\uli{i}\diff\tau\artinput\uli{i, \kappa|k}\nonumber\\
	& + \sum_{\ell=0}^{\kappa-1}e^{\ma{A}\uli{i}((\kappa-\ell-1)\stimes+\delta)}\int_{0}^{\stimes}e^{\ma{A}\uli{i}\tau}\ma{B}\uli{i}\diff\tau\artinput\uli{i, \ell|k},\nonumber\\
	&\mathrm{with~} 0\leq\delta<\stimes\nonumber
\end{align}
\begin{equation}
	\label{eq:inputconst}
	\artinput\uli{i, \mathrm{min}} \leq \artinput\uli{i, \ell|k} \leq \artinput\uli{i, \mathrm{max}}, \forall\ell\in\left\{0,...,h\ul{s}-1\right\}
\end{equation}
\begin{equation}
	\label{eq:stateconst}
	\begin{split}
	\vexset{i}{}_{,\mathrm{min}} \leq \vexset{i}\of{\kappa\stimecons + T|kT} \leq \vexset{i}{}_{,\mathrm{max}}~
	\forall\kappa\in \{0,...,h\ul{b}\}
	\end{split}
\end{equation}
\begin{equation}
	\label{eq:feascond}
	\begin{split}
	\speedset{i}\of{HT+T|kT} = \ve{0},~
	\vey\uli{i}\of{HT+T|kT} = \ve{0}
	\end{split}
\end{equation}
\begin{equation}
	\label{eq:collisionconstr}
	\begin{split}
	\ma{A}\uli{i, j, \mathrm{c}}\begin{bmatrix}
		\posset{i}\of{T|kT}\\
		\posset{i}\of{\stimecol+T|kT}\\
		\vdots\\
		\posset{i}\of{h\ul{c}\stimecol+T|kT}
	\end{bmatrix} \leq \ma{b}\uli{i, j, \mathrm{c}},\\\forall j\in\{1,...,N\},~j\neq i,
\end{split}
\end{equation}
\end{subequations}
with $\tfrac{T}{\stimeobj} = \tfrac{h\ul{o}}{H},\tfrac{T}{\stimecons} = \tfrac{h\ul{b}}{H},\tfrac{T}{\stimecol} = \tfrac{h\ul{c}}{H}\in\mathbb{N}$.
The solution of the optimization problem results in input values of the linear system (\ref{eq:dynamics}) such that a quadratic term between the planned trajectory and the desired target $\vex\uli{i, \mathrm{target}}$, and of the input with positive definite weights $\ma{Q}\uli{i}$, $\ma{R}\uli{i}$ is minimized under several conditions. All conditions are time-discrete and the state is obtained using the solution of linear state space systems (\ref{eq:dynconst}).
The different sample times $\stimes, \stimeobj, \stimecons, \stimecol$ allow us to tune the time density of the constraints independent of the inter-round time $T$, which cannot be tuned freely, because it depends on $T\ul{calc}$ and $T\ul{com}$. The sample times are chosen such that the ends of communication rounds fall on sample time points and the corresponding horizons $h\ul{s}, h\ul{o}, h\ul{b}, h\ul{c}$ are chosen such that all last sample time points fall on the same time $HT$, with prediction horizon $H$. This is important to ensure recursive feasibility (\cite{Park2021}). Because of the delay of calculation and communication, the trajectory starts in the next step $(k+1)T$.
The inputs and states are limited to a minimum and maximum (\ref{eq:inputconst})--(\ref{eq:stateconst}) according to dynamic limits of the UAV. Equation (\ref{eq:feascond}) is important for the feasibility at the next sample time point $(k+1)T$ (\cite{Park2021}). The property $\artinput\uli{i}\of{\tau+T|kT} = \ve{0}$ for all $\tau\geq HT$ then leads to $\vexset{i}\of{\tau+T|kT} = \vexset{i}\of{HT+T|kT}$ for all $\tau\geq HT$. Thus, at the end of the horizon, the UAV stops and stays at its position $\posset{i}\of{HT+T|kT}$.

The last constraint (\ref{eq:collisionconstr}) ensures that the UAV does not collide with others. The matrices $\ma{A}\uli{i, j, \mathrm{c}}$ and $\ma{b}\uli{i, j, \mathrm{c}}$ are calculated using TV-BVC (\cite{TVBVC}), which Figure \ref{fig:constrComp} illustrates.
For every sample time point, the \cu\ calculates the difference vectors between UAV $i$'s and UAV $j$'s last planned trajectory for all $h\in\{0,..., h_\mathrm{c}\}$
 \begin{equation*}
	\begin{split}
\ve{n}\uli{ij}\of{h\stimecol+2 T|(k-1)T} =\ma{\Theta}&{}^{-1}[\posset{j}\of{h\stimecol +2T|(k-1)T} \\&- \posset{i}\of{h\stimecol+2T|(k-1)T}].
	\end{split}
\end{equation*}
Then a plane with the normal vector $\ve{n}\uli{ij}$ is spanned in the center between the UAVs positions. UAV $i$ has to stay on the plane's right side with distance of at least $0.5\hat{r}\ul{min}$
\begin{align}
	\ve{n}\uli{0, ij}\of{h\stimecol+2T|(k-1)T}{}^\transp
\ma{\Theta}&{}^{-1}\left[\posset{j}\of{h\stimecol+2T|(k-1)T}\right.\nonumber\\
&\left. - \posset{i}\of{h\stimecol+T|kT}\right]\nonumber\\ 	\label{eq:lincolconst}
\geq \frac{1}{2}(\hat{r}\ul{min} + ||\ve{n}\uli{ij}(h\stimecol+2T|(&k-1)T)||_2),
\end{align}
where $\ve{n}\uli{0, ij}=\frac{\ve{n}\uli{ij}}{\twonorm{\ve{n}\uli{ij}}}$ and $\hat{r}\ul{min} \geq r\ul{min}$. We will show how to select $\hat{r}\ul{min}$ such that (\ref{eq:col}) holds in Theorem \ref{th:anticollcont}. Combining (\ref{eq:lincolconst}) for all $h\in\{0,...,h\ul{c}\}$, leads to constraint (\ref{eq:collisionconstr}).

After the \cu\ has solved optimization problem (\ref{eq:opt}), it sends the calculated trajectory and inputs to the UAV and all other \cus\ via the wireless network. The UAV sets $\veu\uli{i}\of{\tau+kT} = \veu\uli{i}\of{\tau|kT}$ for $\tau\in[T, 2T)$.
In cases of no recalculation ($\gamma\uli{q i}\of{k}=0, \forall q$), the UAV reuses the previously planned trajectory
similar to \cite{Cai2018}
\begin{align}
	\label{eq:noopttrajplan}\vexset{i}\of{t|kT}= 
		\vexset{i}\of{t+T|(k-1)T}\\
		\label{eq:nooptinputplan}
		\veu\uli{i}\of{t|kT}=\veu\uli{i}\of{t+T|(k-1)T}.
\end{align}
for all $t\geq T$
In both situations, replanning or reusing, 
the resulting trajectories are not closer than $\hat{r}\ul{min}$ at the sampling points.
\begin{lemma}
	\label{lem:colguar}
If $\twonorm{\ve{n}\uli{ij}\of{h\stimecol+2T|(k-1)T}} \geq\hat{r}\ul{min}$ for $i \neq j$ and $\forall h\in\{0,..., h\ul{c}\}$, 
and if (\ref{eq:collisionconstr}) holds for both $i$ and $j$ independent of $\gamma\uli{q i}\of{k}$ and $\gamma\uli{q j}\of{k}~\forall q\in\{1,...,M\}$, then
\begin{align}
	\label{eq:guaranticoll}
	\twonorm{\ma{\Theta}^{-1}\left[\posset{j}\of{h\stimecol+T|kT}-\posset{i}\of{h\stimecol+T|kT}\right]}\geq \hat{r}\ul{min}
\end{align} 
\end{lemma}
\begin{proof}
	If agent $j$'s trajectory is not recalculated ($\gamma\uli{q j}\of{k}=0, \forall q$), it is with $\tau = h\stimecol+T$:
	\begin{align*}
			&\twonorm{\ma{\Theta}^{-1}\left[\posset{j}\of{\tau|kT} - \posset{i}\of{\tau|kT}\right]} \\
			&=\twonorm{\ve{n}\uli{0, ij}\of{\tau+T|(k-1)T}}\\
			&\qquad\times\twonorm{\ma{\Theta}^{-1}\left[\posset{j}\of{\tau|kT} - \posset{i}\of{\tau|kT}\right]}\\ & \geq  \ve{n}\uli{0, ij}\of{\tau+T|(k-1)T}^\transp\ma{\Theta}^{-1}(\posset{j}\of{\tau|kT} - \posset{i}\of{\tau|kT}) \\ &=\ve{n}\uli{0, ij}\of{\tau+T|(k-1)T}^\transp\\
			&\qquad\times\ma{\Theta}^{-1}(\posset{j}\of{\tau+T|(k-1)T} - \posset{i}\of{\tau|kT})\\&\geq \frac{1}{2}(\hat{r}\ul{min} + \twonorm{\ve{n}\uli{ij}\of{\tau+T|(k-1)T}}) \geq\hat{r}\ul{min}.
	\end{align*}
	If $\exists q, \text{~s.t.~} \gamma\uli{q j}\of{k}=1$, we can follow the proof for Lemma 1 of \cite{Zhou2017}. Because (\ref{eq:collisionconstr}) also holds for $j$, we know that (\ref{eq:lincolconst}) is also fulfilled when $i$ and $j$ swap places. Adding it to (\ref{eq:lincolconst}), we get (using $\ve{n}\uli{ij} = -\ve{n}\uli{ji}$)
		\begin{align}
			&\ve{n}\uli{0, ij}\of{\tau+T|(k-1)T}^\transp\ma{\Theta}^{-1}\left[(\posset{j}\of{\tau+T|(k-1)T}\right. \nonumber\\&\qquad\left. - \posset{i}\of{\tau|kT})- (\posset{i}\of{\tau+T|(k-1)T} - \posset{j}\of{\tau|kT})\right]\nonumber\\ & = 
		\ve{n}\uli{0, ij}\of{\tau+T|(k-1)T}^\transp\ma{\Theta}^{-1}(\posset{j}\of{\tau|kT} - \posset{i}\of{\tau|kT}) \nonumber\\& \qquad+\twonorm{\ve{n}\uli{ij}\of{\tau+T|(k-1)T}}\nonumber\\
		& \geq 
			\hat{r}\ul{min} + \twonorm{\ve{n}\uli{ij}\of{\tau+T|(k-1)T}}.\nonumber
		\end{align}
	Canceling $\twonorm{\ve{n}\uli{ij}\of{\tau+T|(k-1)T}}$ on both sides and applying the Cauchy-Schwarz inequality to the right side, we get (\ref{eq:guaranticoll}). 
\end{proof}
With this result, we can state:
\begin{theorem}
	\label{th:feas}
	If $\vexset{i}\of{t+T|(k-1)T}$ fulfills (\ref{eq:dynconst})--(\ref{eq:collisionconstr}) and if\\ $||\ve{n}\uli{ij}(h\stimecol+T|(k-1)T)||_2 \geq\hat{r}\ul{min}$ for all pairwise different UAVs $i\neq j$ and $\forall h\in\{0,...,h\ul{c}\}$, then the subsequent trajectory $\vexset{i}\of{t|kT}$ generated by (\ref{eq:noopttrajplan}) also fulfills the constraints, the optimization problem (\ref{eq:opt}) is feasible at the following timestep $kT$ and $\twonorm{\ve{n}\uli{ij}\of{h\stimecol+T|kT}} \geq\hat{r}\ul{min}$.
\end{theorem}
\begin{proof}
	We use techniques from the proof of Theorem 2 in \cite{Park2021}.
	Because trajectory $\vexset{i}\of{t|kT}$ generated by (\ref{eq:noopttrajplan}) is equal to $\vexset{i}\of{t+T|(k-1)T}$ and the time duration $T$ is an integer multiple of the constraint sample times, it fulfills (\ref{eq:dynconst})--(\ref{eq:feascond}) in the first part ($t\leq HT$). The second part of the trajectory for $t>HT$ fulfills those constraints too, because the UAV stays at its position. 
	
%
The positions $\posset{i}\of{t|kT}$ and $\posset{i}\of{t+T|(k-1)T}$ are also equal with the same argumentation as above. If we now set this into the left side of (\ref{eq:lincolconst}), the left side is equal to
 $||\ve{n}\uli{ij}(h\stimecol+2T|(k-1)T)||_2$, which is bigger than the right side of (\ref{eq:lincolconst}) $\forall h\in\{0,...,h\ul{c}\}$ due to the second condition of this theorem and because the UAV stops for $t\geq HT$. Thus, the constraint (\ref{eq:collisionconstr}) holds. Because (\ref{eq:noopttrajplan}) fulfills all constraints of (\ref{eq:opt}), (\ref{eq:opt}) is feasible at $kT$. The argumentation above also holds, if $i$ and $j$ swap places. Lemma \ref{lem:colguar} then states that $\twonorm{\ve{n}\uli{ij}\of{h\stimecol+T|kT}} \geq\hat{r}\ul{min}$.
%
\end{proof}

It directly follows from the recursive nature of this theorem that as long as the initial trajectories fulfill the conditions, we can generate trajectories without any collisions at sample time points with sample time $\stimecol$ using (\ref{eq:opt}) or (\ref{eq:noopttrajplan}) at all times.
%
However, this does not guarantee that (\ref{eq:col}) is fulfilled and the UAVs do not collide. The trajectories might get too close in between the sample time points (\cite{Luis2019}). The following theorem states under which conditions the swarm is guaranteed to be collision free:
\begin{theorem} \label{th:anticollcont}
	If $\vex\uli{i}\of{t|-T}$ fulfills the conditions of Theorem (\ref{th:feas}), if $\vex\uli{i}\of{T|-T}=\vex(0)$,
	and if $\hat{r}\ul{min} - \dpmax{i}{\stimecol} -\dpmax{j}{\stimecol} \geq r\ul{min}$~$\forall i,j\in\{1,...,N\}, i\neq j$ with
		\begin{align}
			\nonumber
			\Delta\ve{p}\uli{i, \mathrm{max}}(\tilde{T})&= \max_{\vexset{i}\in[\vexset{i}{}_{,\mathrm{min}},\vexset{i}{}_{,\mathrm{max}}], \artinput\uli{i}\of{t}\in[\artinput\uli{i,\mathrm{min}},\artinput\uli{i,\mathrm{max}}], \tau\in\left[0;\tilde{T}\right]}\bigg[\\ ||\begin{bmatrix}
					\ma{\Theta} & \ma{0}
				\end{bmatrix} \big[(e&{}^{\ma{A}\uli{i}\tau}-1)\vexset{i} +\int_{0}^{\tau}e^{\ma{A}\uli{i}\tilde{\tau}}\ma{B}\artinput\uli{i}(\tau-\tilde{\tau})\diff\tilde{\tau}\big]||_2\bigg]\nonumber
		\end{align}
	as upper bounds of the maximum distance, the position can travel between two sample time points with a sample time $\tilde{T}$, then the UAV swarm fulfills (\ref{eq:col}). 
\end{theorem}
\begin{proof}
	Feasible solutions of the trajectory planning exist for every $k\geq0$ (Theorem \ref{th:feas}).
	 We select $k, h\ul{c}\in\mathbb{N}$ and $0\leq\delta< \stimecol$ such that $h\ul{c}\stimecol<T$ and $t=kT+\mathcal{T}$ with $\mathcal{T}=h\ul{c}\stimecol+T+\delta$.
	  Using Theorem \ref{th:feas} we know that
	  \begin{equation*}
	  	\begin{split} &\twonorm{\ma{\Theta}^{-1}[\posset{i}\of{t}-\posset{j}\of{t}]}=\twonorm{\ma{\Theta}^{-1}[\posset{i}\of{\mathcal{T}|kT}-\posset{j}\of{\mathcal{T}|kT}]}\\
	  	&\geq\twonorm{\ma{\Theta}^{-1}[\posset{i}\of{h\ul{c}\stimecol+T|kT}-\posset{j}\of{h\ul{c}\stimecol+T|kT}]}\\&\qquad-\dpmax{i}{\stimecol}-\dpmax{j}{\stimecol}\\
	  	&\geq \hat{r}\ul{min}-\dpmax{i}{\stimecol}-\dpmax{j}{\stimecol}.
	  \end{split}
	\end{equation*}
\end{proof}
Theorem \ref{th:anticollcont} is independent of the chosen ET. 
There are two reasons for this independence. First, the swarm must have collision-free initial trajectories for Theorem \ref{th:anticollcont} to hold. Second, each UAV can recursively generate collision-free trajectories with (\ref{eq:noopttrajplan}) without triggering. However, when a UAV's replanning is not triggered for $H$ consecutive times, it will stop and stay at a position which might not be its target position. The choice of the ET is thus important to quickly steer the UAVs into their targets.

\textbf{Priority based trigger rule.} As a possible trigger rule, we chose  PBT (\cite{PredictiveTriggering}). With PBT, every UAV gets assigned a priority $g\uli{i}\of{kT}$. Because all \cus\ have the same information due to the many-to-all communication architecture, they can calculate the priorities of all UAVs locally without any additional communication. \cu\ $q$ then replans the trajectory of the UAV with $q$-th highest priority. The choice of the priority and its parameters allow us to tune the selection of UAVs and thus the performance of the overall system. We chose the following priority:
\begin{equation}
	\label{eq:prio}
	\begin{split}
		g\uli{i}\of{kT} =& \alpha\uli{1}\twonorm{\ve{d}\uli{i, \text{target}}}+\alpha\uli{2}\Delta t\uli{i}\of{kT}\\- \alpha\uli{3}&\max\Big(\cos\of{\beta},\sum_{j\neq i}\xi\frac{\ve{d}\uli{i, \text{target}}^\transp\ve{d}\uli{ij}}{||\ve{d}\uli{i, \text{target}}||_2||\ve{d}\uli{ij}||_2}\Big),
	\end{split}
\end{equation}
where $\ve{d}\uli{ij} = \posset{j}\of{2T|(k-1)T}-\posset{i}\of{2T|(k-1)T}$, $\ve{d}\uli{i, \text{target}} = \pos\uli{i, \text{target}}-\posset{i}\of{2T|(k-1)T}$ and
$\xi=\max\left(0, ||\ve{d}\uli{i, \text{target}}||_2 \right.$ $- ||\left. \ve{d}\uli{ij}||_2\right)$. $\Delta t\uli{i}\of{kT}$ describes how long the trajectory of UAV $i$ was not replanned. $\alpha\uli{1}, \alpha\uli{2}, \alpha\uli{3}\geq 0$ are weights of the summands. The last term measures how many UAVs lie in a cone with angle $\pm\beta$ between the UAV $i$ and its target and how close they are (factor $\xi$). The more UAVs, the more likely $i$ will get blocked by others and thus replanning is ineffective.

\section{Experiments and Discussion}
\pgfplotsset{compat=1.5.1}
\begin{figure}
	\centering
	\begin{tikzpicture}[spy using outlines={rectangle, magnification=2.5,connect spies}]
		\begin{axis}[xlabel={$t$ in \SI{}{\second}}, ylabel={Mean distance to target in \SI{}{\meter}}, ylabel style={text width=2.5cm}, xmin=0, xmax=60, legend style={at={(0.98,0.98)},anchor=north east ,draw=black,fill=white,align=left, nodes={scale=0.5, transform shape}}, height=0.2\textwidth, width=0.45\textwidth]
			\addplot [color=blue] table [x index = {0}, y index = {1}]  {plotData/RoundRobinFinal2_15.csv};
			\addplot [color=red] table [x index = {0}, y index = {1}]  {plotData/RoundRobinFinal2_20.csv};
			\addplot [color=black!30!green] table [x index = {0}, y index = {1}]  {plotData/RoundRobinFinal2_25.csv};
			\addplot [color=cyan] table [x index = {0}, y index = {1}]  {plotData/NoETFinal_25.csv};
			\addplot [color=blue, dashed] table [x index = {0}, y index = {1}]  {plotData/EventTriggeringFinal2_15.csv};
			\addplot [color=red, dashed] table [x index = {0}, y index = {1}]  {plotData/EventTriggeringFinal2_20.csv};
			\addplot [color=black!30!green, dashed] table [x index = {0}, y index = {1}]  {plotData/EventTriggeringFinal2_25.csv};
			
			\coordinate (spypoint) at (axis cs:25,0.2);
			\coordinate (spyviewer) at (axis cs:37,3);
			\spy[width=5cm,height=1cm] on (spypoint) in node [fill=white] at (spyviewer);
		\end{axis}
	\end{tikzpicture}

	\caption{Mean distance to target. \capt{Dashed lines show the results of ET-DMPC with PBT, while the solid lines represent the results of ET-DMPC with round-robin scheduler. Blue: $N=15, M=10$, Red: $N=20, M=10$, Green: $N=25, M=10$, Cyan: $N=25, M=25$. In the mean, PBT is faster and more UAVs reach their target (magnification).}}
	\label{fig:successrate}
\end{figure}
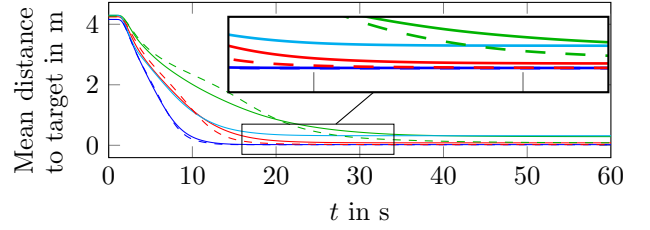
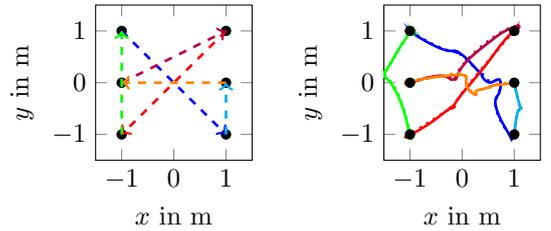
\begin{figure}
	\newcommand{\figwidth}{0.20\textwidth}
	\centering
	\begin{tikzpicture}
		\newcommand{\radiusplot}{0.1}
		\newcommand{\plotlinewidth}{1}
		\begin{axis}[name=plot1, xlabel={$x$ in \SI{}{\meter}}, ylabel={$y$ in \SI{}{\meter}}, xmin=-1.5, ymin=-1.5, xmax=1.5, ymax=1.5 ,legend style={at={(0.98,0.021)},anchor=south east ,draw=black,fill=white,align=left, nodes={scale=0.7, transform shape}}, height=\figwidth, width=\figwidth]
			\fill (axis cs:1, 1) circle [black, fill,radius=\radiusplot];
			\fill (axis cs:-1, 1) circle [black, fill,radius=\radiusplot];
			\fill (axis cs:-1, -1) circle [black, fill,radius=\radiusplot];
			\fill (axis cs:1, -1) circle [black, fill,radius=\radiusplot];
			\fill (axis cs:-1, 0) circle [black, fill,radius=\radiusplot];
			\fill (axis cs:1, 0) circle [black, fill,radius=\radiusplot];
			
			\draw[->, dashed, blue, line width=1](axis cs:-1, 1)--(axis cs:1, -1);
			\draw[->, dashed, red, line width=1](axis cs:1, 1)--(axis cs:-1, -1);
			
			\draw[->, dashed, purple, line width=1](axis cs:-1, 0)--(axis cs:1, 1);
			\draw[->, dashed, orange, line width=1](axis cs:1, 0)--(axis cs:-1, 0);
			
			\draw[->, dashed, green, line width=1](axis cs:-1, -1)--(axis cs:-1, 1);
			\draw[->, dashed, cyan, line width=1](axis cs:1, -1)--(axis cs:1, 0);
		\end{axis}
		\coordinate[right=0.5cm of plot1] (coord);
		\begin{axis}[name=plot2, at=(coord), anchor=left of west, xlabel={$x$ in \SI{}{\meter}}, ylabel={$y$ in \SI{}{\meter}}, xmin=-1.5, ymin=-1.5, xmax=1.5, ymax=1.5 ,legend style={at={(0.98,0.02)},anchor=south east ,draw=black,fill=white,align=left, nodes={scale=0.7, transform shape}}, height=\figwidth, width=\figwidth]
\addplot [color=blue, line width=\plotlinewidth] table [x index = {0}, y index = {1}]  {plotData/ExperimentPosition.csv};
\addplot [color=blue, line width=\plotlinewidth, dotted] table [x index = {0}, y index = {1}]  {plotData/ExperimentTargetPosition.csv};
\addplot [color=red, line width=\plotlinewidth] table [x index = {3}, y index = {4}]  {plotData/ExperimentPosition.csv};
\addplot [color=red, line width=\plotlinewidth, dotted] table [x index = {3}, y index = {4}]  {plotData/ExperimentTargetPosition.csv};
\addplot [color=purple, line width=\plotlinewidth] table [x index = {6}, y index = {7}]  {plotData/ExperimentPosition.csv};
\addplot [color=purple, line width=\plotlinewidth, dotted] table [x index = {6}, y index = {7}]  {plotData/ExperimentTargetPosition.csv};
\addplot [color=orange, line width=\plotlinewidth] table [x index = {9}, y index = {10}]  {plotData/ExperimentPosition.csv};
\addplot [color=orange, line width=\plotlinewidth, dotted] table [x index = {9}, y index = {10}]  {plotData/ExperimentTargetPosition.csv};
\addplot [color=green, line width=\plotlinewidth] table [x index = {12}, y index = {13}]  {plotData/ExperimentPosition.csv};
\addplot [color=green, line width=\plotlinewidth, dotted] table [x index = {12}, y index = {13}]  {plotData/ExperimentTargetPosition.csv};
\addplot [color=cyan, line width=\plotlinewidth] table [x index = {15}, y index = {16}]  {plotData/ExperimentPosition.csv};
\addplot [color=cyan, line width=\plotlinewidth, dotted] table [x index = {15}, y index = {16}]  {plotData/ExperimentTargetPosition.csv};
\fill (axis cs:1, 1) circle [black, fill,radius=\radiusplot];
\fill (axis cs:-1, 1) circle [black, fill,radius=\radiusplot];
\fill (axis cs:-1, -1) circle [black, fill,radius=\radiusplot];
\fill (axis cs:1, -1) circle [black, fill,radius=\radiusplot];
\fill (axis cs:-1, 0) circle [black, fill,radius=\radiusplot];
\fill (axis cs:1, 0) circle [black, fill,radius=\radiusplot];
		\end{axis}
	\end{tikzpicture}
	\caption{Changing positions. \capt{The left plot shows which UAV has to fly to which position. The algorithm is able to lead all UAVs to their target position (right plot). Dotted lines are planned, solid lines real trajectories.}}
	\label{fig:circlechange}
\end{figure}
{\bf Simulation environment.} To evaluate the proposed methods, we have built a simulation of the system\footnote{The code with all parameters can be found in \url{https://github.com/Data-Science-in-Mechanical-Engineering/ET-distributed-UAV-path-planner}.}. It simulates a swarm of Crazyflies using gym-pybullet-drones
(\cite{panerati2021learning}). We approximate the quadcopter dynamics as a triple integrator for the DMPC (\cite{Wang2021}). Because it does not represent the real quadcopter dynamics accurately, the predicted positions using the linear model are set as setpoints of local position controllers of the UAVs (\cite{Luis2019}). We chose a sampling time of $T=\SI{333.33}{\milli\second}$ ($T\ul{calc}=\SI{233.33}{\milli\second}$, $T\ul{com}=\SI{100}{\milli\second}$). We set the minimal distance $\hat{r}\ul{min}=\SI{0.7}{\meter}$ and $\theta=\diag[1, 1, 2]$. 
The space where the UAVs can fly is limited to a cuboid of $\SI{5}{\meter}\times\SI{5}{\meter}\times\SI{5}{\meter}$. 
Initial and target positions are generated randomly using a truncated uniform random distribution such that initial/target positions scaled with $\ma{\theta}$ are not closer to each other than $\hat{r}\ul{min}$. We call one realization of initial and target positions scenario in the following. We could fit up to 25 UAVs in the space.


{\bf Influence of ET.} To determine the influence of ET, we compare PBT to a round-robin trigger. Figure \ref{fig:successrate} shows experiments with 15, 20 and 25 UAVs for $M=10$ \cus\ and TV-BVC with 1000 scenarios for each setup. In all setups, PBT steered the UAVs faster to their targets than round-robin. Additionally, the number of UAVs reaching their target is higher for PBT (\SI{98.5}{\percent}) than round-robin (\SI{92.7}{\percent}). But PBT cannot avoid completely that some UAVs are not able to reach their target.
This phenomenon is called deadlock
and caused by the absence of a centralized coordinator (\cite{Luis2019, Park2021}). Every UAV's trajectory is just optimal regarding the corresponding UAV. It thus would not make room for another UAV if this trajectory moved away from the target position. We noticed that some deadlocks lead to infeasibility due to numerical issues of the solver, which also hinders to dissolve them. Soft constraints could help to avoid this issue (\cite{Luis2019}). To avoid deadlocks completely, the targets $\pos\uli{i, \mathrm{target}}$ itself can be changed using a cooperative planner (\cite{Park2021}).

Furthermore, we compared the performance of a swarm with an equal number of UAVs and \cus\ ($N=M=25$) to one with less \cus\ than UAVs ($M=10<N=25$). We noticed that although some UAVs reached their target faster, there were a lot of deadlocks for $M=25$. This leads to the shape of the mean distance to target in Figure \ref{fig:successrate}. 
Thus, in this scenario, ET not only saves \SI{60}{\percent} of bandwidth and computing hardware, it also improves the performance in the mean.

{\bf Hardware-in-the-loop experiments.} To validate the sufficiency of the choice of a linear model for quadcopter trajectory generation, we tested our method using a hardware-in-the-loop (HiL) simulation with six real Crazyflies flying indoor. We  simulated the wireless network and one \cu\ on a local PC, which runs Crazyswarm (\cite{Crazyswarm}) and is connected to a motion capture system to obtain the positions of the crazyflies.
Figure \ref{fig:circlechange} shows an example trajectory of our HiL-simulation, where UAVs swap positions. We observed no dangerous flight maneuver and unstable behavior of the quadcopters when following the trajectory as can be seen in \url{https://youtu.be/2UzqOnUJQCA}. The choice of a linear model in our ET-DMPC seems to be sufficient for real quadcopters.

\section{Conclusions}
This paper presented an event-triggered path-planning algorithm for UAV swarms. The developed computation scheme selects a subset of UAVs and replans the trajectories of only these on stationary computation units. The DMPC based path planner guarantees recursive feasibility and collision free trajectories for linear UAV dynamics. Our approach has been demonstrated to reliably steer UAVs without any collisions.
This shows that event-triggering can be used to lower the communications and computational resources needed for path planning while guaranteeing collision free trajectories. 
In future work, we will add static obstacle avoidance to ET-DMPC using a safe flight corridor (\cite{Park2020, Park2021}). Additionally, we will investigate distributed cooperative planners to avoid deadlocks.

\section*{Acknowledgements}
We thank Sebastian Giedyk for his help with the development and implementation of the quadcopter testbed, Alexander von Rohr and Kurt Capellmann for their help with the execution of the simulations, and Christian Fiedler and Henrik Hose for feedback on the manuscript.

\bibliography{references}             
                                                   






\end{document}